\documentclass[aps,twocolumn,floatfix,showpacs]{revtex4}
\usepackage{graphicx}
\usepackage{graphics}
\newcommand{\be}{\begin{equation}}
\newcommand{\ee}{\end{equation}}
\newcommand{\bea}{\begin{eqnarray}}
\newcommand{\eea}{\end{eqnarray}}

\begin{document}


\title{Random Walks with Shrinking Steps: \\ First Passage Characteristics}


\author{Tongu\c{c} Rador}
\email[]{tonguc.rador@boun.edu.tr}
\affiliation{Bo\~{g}azi\c{c}i University \\ Deparment of Physics \\ Bebek 34342 , \.Istanbul, Turkey}
\author{Sencer Taneri}
\email[]{taneri@gursey.gov.tr}
\affiliation{Feza G\"ursey Institute \\ Emek. Mah. No:68 \\ \c{C}engelk\"oy 34864 , \.Istanbul, Turkey}


\date{\today}

\begin{abstract}
We study the mean first passage time of a 
one-dimensional random walker with step sizes decaying exponentially in discrete 
time. That is step sizes go like $\lambda^{n}$ with $\lambda\leq1$ . We also 
present, for pedagogical purposes, a continuum system with a diffusion constant 
decaying exponentially in continuous time. Qualitatively both systems are alike 
in their global properties. However, the discrete case shows very rich mathematical 
structure, depending on the value of the shrinking parameter, such as self-repetitive 
and fractal-like structure for the first passage characteristics. The results we 
present show that the most important quantitative behavior of the discrete case 
is that the support of the distribution function evolves in time in a rather 
complicated way in contrast to the time independent lattice structure of the 
ordinary random walker. We also show that there are critical values of $\lambda$ 
defined by the equation $\lambda^{K}+2\lambda^{P}-2=0$ with $\{K,N\}\in{\mathcal N}$ 
where the mean first passage time undergo transitions.

\end{abstract}


\maketitle

\section{Introduction}

The common model of random walk is an integral 
part of almost all scientific disciplines in such a way that it 
would not be an overstatement to call it
a meta-model. In this respect it is important to 
look for extensions and possible generalizations of the model. 
In fact many of these extensions are already present: the
ordinary random walk is solvable in many lattices with 
arbitrary nearest neighbor sites,
the continuum limits are known for various geometries and 
generalizations for space dependent diffusion constant 
also exist in nature.

Another likely extension of a random walk could be the case of a 
walker with step sizes shrinking with time. Although there is 
considerable work \cite{art5}-\cite{art10} on this in mathematical literature,
the idea has not been studied extensively from a physics perspective:
see \cite{art1}-\cite{art4}. 
However there is a rather straightforward motivation for 
it to be important. The idea is the dichotomy between a 
Brownian particle and an {\em active} random walker. In the 
former case the particle is merely pushed around
by the surrounding molecules. That is, it is under the 
influence of a random external force. In the latter case however, there is a
possibility for internal energy dissipation and hence
the step sizes (the ability to diffuse in the medium) might decrease 
in time. It is a difficult exercise to visualize
how this might occur in atomic scales. 
It is more likely that this motivation would make sense only in some form
of macroscopic mean field theory.

If we can be allowed to be a little humorous, 
the very analogy of a {\em drunk man} for the ordinary random walk 
is in essence closer to a walker with shrinking step sizes. This example might 
at first seem rather non-scientific, but it brings in mind how
biological systems, with rather short memory, propagate in nature 
without energy intake, let it be a fish in a bowl or a fly
in a room. It is most likely that step size changes in nature are not 
fierce under normal circumstances so their effect is immaterial. However, under 
extreme circumstances such as lions hunting prey the internal energy 
dissipation could be important and then it is a matter of determining how 
the step sizes change in time from biological principles.

Nevertheless, the aim of this paper is not to make a full case for 
the physical relevance of a random walker with shrinking step sizes. We
believe that random walk would remain a meta-model with this extension as well.

The motivation for this work is the very interesting first passage characteristics of the
ordinary random walk, such as the dichotomy between {\em certain passage} from any point and
{\em infinite mean first passage time} from that point for a random walker in one dimension. 
Various processes in nature proceed by  first passage processes. For example the hunt will be over when the prey and predator meets for the first time, or a neuron will fire when the electric potential first reaches
a threshold value, etc. For further information on the study of first passage time for various systems we refer the reader to \cite{art11} (and references therein) for a review and introductory book and to \cite{art12}-\cite{art16}  for recent literature.

The outline of the paper is as follows. In Section~\ref{sec2}, for purposes of
motivation and developping a qualitative understanding of the problem, we
consider the first passage characteristics of a diffusion equation with a
diffusion constant that decreases exponentially with time. In particular we emphasize the asymptotics of
the mean first passage time for later comparison with the discrete case.

Section~\ref{sec3} introduces the discrete random walker with shrinking step sizes with an emphasis on the most general aspects of the model. In particular we emphasize the fact that the support of the distribution evolves in time in a complicated way rendering an analytical study almost impossible.

The first passage characteristics of the discrete random walker with exponentially decreasing step sizes are presented in sections \ref{sec4} and \ref{sec5}. In \ref{sec4} the model is  studied numerically  for various parameters and  both local and global properties of the mean first passage time are analyzed. In \ref{sec5} we present an in depth analysis of the global properties. In these sections we also contrast  the results to the continuum model and substantiate that the two models agree on their general behaviors if not in every detail.


\section{A continuum example}\label{sec2}

For presentation purposes, before getting into the intricacies of a discrete 
random walk, we expose the behavior of a shrinking diffusion constant in 
continuum diffusion equation. That is, we consider the following,

\be\label{eq:diff1}
\frac{\partial P}{\partial t}=D\;e^{-t/\tau}\;\frac{\partial^{2} P}{\partial x^{2}}\;\;.
\ee

We would like to solve this in a semi-infinite line with the initial 
condition that $P(x,0)=\delta(x)$ and absorbing boundary conditions 
at $x=x_{o}$, that is, $P(x_{o},t)=0$. 

The differential equation in (\ref{eq:diff1}) can be easily solved with the above conditions to give,

\bea
P(x,T)&=&\frac{1}{\sqrt{4\pi D T}}\left[e^{-\frac{x^{2}}{4DT}}-e^{-\frac{(x-2x_{o})^{2}}{4DT}}\right]\;\;, \\
T&=&\tau\;\left(1-e^{-t/\tau}\right)\;\;.
\eea

The first passage probability to $x_{o}$ is just the flux to this point. The flux can
be read from the diffusion equation if one writes it as 
$\partial P/\partial t=-\partial F/\partial x$. This will give

\be
F(x_{o},t)=\frac{x_{o}}{\sqrt{4\pi D T^{3}}}\exp\left[\frac{x_{o}^{2}}{4 D T}\right]\;\;.
\ee

\noindent From this the zeroth moment or the eventual first passing 
probability can be calculated to give

\be
F_{0}(x_{o})=1-{\rm Erf}\left[\frac{x_{o}}{\sqrt{4 D \tau}}\right]\;\;.
\ee

\noindent Here, ${\rm Erf}(u)=\frac{2}{\sqrt{\pi}}\int_{0}^{u} dt e^{-t^{2}}$ 
is the error function. The result above should be contrasted to the ordinary 
diffusion equation where $F_{0}(x_{o})=1$. In the problem under consideration there
 is a finite first passage probability from $x_{o}$, however this is rapidly vanishing with 
increasing $x_{o}$ as one would expect. 

The survival probability, that is the probability to never have been 
to $x_{o}$ at time $t$, is defined as

\be
S(x_{o},t)\equiv 1-\int_{0}^{t} F(x_{o},t') dt'\;\;,
\ee

\noindent which gives an exact result

\be
S(x_{o},t)={\rm Erf}\left[ \frac{x_{o}}{\sqrt{4 D \tau\left(1-e^{-t/\tau}\right)}} \right]\;\;.
\ee

\noindent This predicts a finite survival probability 
$S(x_{o},\infty)={\rm Erf}(\frac{x_{o}}{\sqrt{4D\tau}})$ for infinite time, again in
contrast to the ordinary random walk with the survival probability 
vanishing as $\frac{x_{o}}{\sqrt{4Dt}}$ as $t\to\infty$.

The mean first passage time can also be calculated. However in this case the result
is only available as a series,

\be\label{first1}
\left<t(x_{0)}\right>=\frac{\tau}{1-{\rm Erf}(u)} \sum_{n=1}^{\infty} \frac{1}{n}\;\;. u^{2n}\Gamma(0.5-n,u^{2})\;,
\ee

\noindent With $u\equiv\frac{x_{o}}{\sqrt{4D\tau}}$ and 
$\Gamma(a,z)\equiv\int_{z}^{\infty}dt\;e^{-t}t^{z-1}$, the 
incomplete gamma function. To understand the behavior of this quantity 
we study its asymptotic behavior. The analytic properties of the error 
function and the incomplete gamma function are readily available in textbooks. 
The results for the two regimes of interest are,

For $u\to 0$ 

\be\label{firstas}
\left<t(x_{o})\right> \simeq \sqrt{\frac{\tau}{\pi}} \frac{x_{o}}{\sqrt{D}}-\frac{\pi-2}{2\pi}\frac{x_{o}^{2}}{D}+\ldots
\ee

For $u\to\infty$

\be\label{firstas2}
\left<t(x_{o})\right>\to \tau \sum_{n=1}^{\infty} \frac{1}{n}\;\;-\tau^{2}\frac{4D}{x_{o}^{2}}\sum_{n=1}^{\infty}1+\ldots\;\;.
\ee

\noindent So as $u\to\infty$ we expect a logarithmic divergence with a 
linear dependence on $\tau$. 

We are now equipped with enough intuition to attack the more 
difficult case of a discrete random walk with shrinking steps.

\section{Random walks with shrinking steps}\label{sec3}

Consider a random walker which hops equally likely to left and 
right by some distance $s_{n}$ at a given time $n$. Then the 
random variable for the position of the walker is
simply

\be{\label{eq:1}}
x(N)=\sum_{n=1}^{N} \epsilon_{n}\;s_{n} \;\; .
\ee

\noindent with $\epsilon_{n}=\pm 1$ the random variable for hopping. 
The mean of $x(N)$ , as expected, is zero: the random walker does not 
propagate in the mean but rather in the standard deviation which is given by

\be{\label{eq:2}}
\left< x^{2}(N)\right> =\sum_{n=1}^{N} \; s^{2}_{n} \;\; .
\ee

From this relation it follows that if the sum converges 
to a constant as $N\to\infty$ there can be no continuum limit \footnote{This argument is presented also in \cite{art2}.} . To clarify the reasoning let us
remember how one takes the continuum limit. We first introduce a lattice spacing $\delta x$, which
would mean that the variance becomes

\[
\left< x^{2}(N)\right> =\delta x^{2} \sum_{n=1}^{N} \; s^{2}_{n} \;\; .
\]

Now if there is a continuum limit we should be able to take  $\delta x\to 0$ with $N\to\infty$ and
still get a nonzero standard deviation. For example in the ordinary random walk  we have $s_{n}=1$, 
which in turn means that $\left< x^{2}(N)\right>=\delta x^{2} N$. So letting $N\to\infty$ can be done by $N\equiv t/\delta t$ and $\delta t\to 0$. The limit would make sense if  $\delta x^{2}/\delta t$ remains
a constant upon taking the limits. Actually this is how one defines the diffusion constant. On the other
hand if the sum in (\ref{eq:2}) converges to a constant independent of $N$ as $N\to\infty$,  we get zero standard deviation upon taking $\delta x\to 0$. Such a behavior will be present for example if $s_{n}=\lambda^{n}$ with $\lambda<1$: the model which is under consideration in this paper.

This pathology is present because the support of the probability distribution function 
for the random variable $x(N)$ evolves in time, in sharp contrast to 
the ordinary random walk for which the walker always resides 
on a lattice site: a time independent regular structure allowing a limit for zero lattice spacing.

Disregarding this difficulty one might be tempted to assume that the use of known machinery such as
Fourier transforms will be useful. However this does not generally help in practice unless special cases occur, such as infinite time limit or a particular shrinking parameter. The fact that the support of the distribution evolves in time in a generally non-uniform way will ultimately infest these approaches, especially if one is interested in the temporal evolution in the finite time regime. 

For example in \cite{art1} the case $s_{n}=\lambda^{n}$ has been shown to be exactly solvable for $\lambda=1/2$. This case is exactly solvable since at each time step the support of the distribution is

\bea{\label{eq:6}}
N=1 && \{-1,1\} \nonumber \\
N=2 && \{-3/2,-1/2,1/2,3/2\} \nonumber \\
N=3 && \{-7/4,-5/4,-3/4,-1/4,1/4,3/4,5/4,7/4\} \nonumber \\
\vdots && \vdots \nonumber
\eea

The support evolves, but at each time step it is a regular, equally-spaced lattice albeit with different lattice spacings for different times. For slightly more general cases, namely the infinite time limit for $\lambda=1/2^{m}$, 
studied in \cite{art1}, what happens is that the supports are actually unions of $m$ evenly 
distributed support sets which results
in mathematical terms to products (convolutions) of the distributions 
in momentum (position) space. This is the reason why these cases are easily manageable. 

\section{Geometric Shrinking: Mean First Passage Time}{\label{sec4}}

In this section we will investigate the mean first passage time 
of a random walker with $s_{n}=\lambda^{n}$. Since there are 
no generally tractable analytic methods we will resort to numerical 
methods in this section. Contrary to the ordinary random walk, yielding 
infinite mean first passage time and hence rendering a direct Monte-Carlo analysis rather hard if not impossible \footnote{We mean a direct simulation would be hard in the ordinary random walk for large times. One can in principle study the first passage time by other numerical methods such as measuring the flux to a point. Such an alternate approach would be more feasible than a direct analysis. }, the walker we are interested in ultimately comes to a stop. So a  
numerical analysis is indeed very feasible. There are however certain issues 
to be resolved in the present case that do not exist in the ordinary random walk. We start this chapter
by listing the numerical and simulational peculiarities we have encountered when studying the system. We then
work out the exactly soluble case of $\lambda\leq1/2$ as a first exercise. Afterwards we present a case study for $\lambda=0.55$ where we lay out the results for the mean first passage time along with  digressions on its various aspects. We
have also present simulations for $\lambda=\sqrt{2}$ and $\lambda=0.9$ in order to better convey
the ideas and to have  contrasts  to the case $\lambda=0.55$ and $\lambda\leq1/2$.

\subsection{Numerical Peculiarities}
\noindent{\bf First passage from a point:} The fact that the support 
of the distribution evolves haunts us here again. In the ordinary random 
walk a first passage from a point
is actually {\em to occupy} that point for the first time. In the case of a walker 
with shrinking steps it is no longer possible 
to predict that a given point is part of the support set at a given time. 
However the idea of {\em passing} certainly makes sense: that
is to be {\em to the right of a point for the first time} is a 
meaningful statement. This unfortunately has the corollary that {
\em a particular path may contribute to the first passage distribution 
of more than one point if these points are close}. As we will see, due to its local aspect, this effect
will not alter the global properties of the system.

\noindent{\bf Falling out of range:} In the ordinary random walk the walker 
never falls out of range of a point. But, with the walker we are interested in  
this will be possible. Certain
walks will fall out of range of a point due to a {\em bad choice of direction} 
at some point earlier in time. One could say earlier mistakes are punished
more severely. However in terms of computer simulations 
this is more than welcome since it
eliminates a walk that would never reach a point. 
This is a great help in reducing the simulation time.

\noindent{\bf Spatial Resolution:} A random walker with 
geometrically shrinking step size will theoretically walk for an 
infinite amount of time. However in practice  one can assume that after 
some time it will not move appreciably and the subsequent walk will have negligible effect.
We have always chosen such a cut-off time such that the subsequent step sizes would be less
then $10^{-9}$.

With these in mind we have performed several simulation as follows. 
We start the walker at
the origin and make a histogram of the time values for which the 
walker first passes a point $x_{o}$. This defines the first passage 
probability distribution $F(x_{o},N)$. We then
compute the normalized mean first passage time as follows

\be{\label{eq:7}}
\left< t(x_{o}) \right> \equiv \frac{F_{1}(x_{o})}{F_{0}(x_{o})}\;\;,
\ee

\noindent with 

\be{\label{eq:8}}
F_{n}(x_{o}) \equiv \sum_{N=0}^{\infty} N^{n} F(x_{o},N)\;\;.
\ee

\noindent defining the moments of $F(x_{o},N)$.

The general details of the simulations were as follows. For a given $x_{o}$, 
we let the walker walk $10^{5}$ times upto the step number cutoff defined so 
that the spatial resolution is below $10^{-9}$. After this bunch is completed we have a single
distribution for $F(x_{o},N)$. We repeat this procedure about 200 times
to get a statistical histogram  from which  everything can be
computed. The procedure is repeated for different values of $x_{o}$. 

\subsection{Exact result for $\lambda\leq1/2$}

This case has a very simplifying aspect: as time passes the walker always falls out of range of any point it occupied in the past. Another way of stating this fact is that the support of the distribution is a Cantor
Set which never crosses itself. Therefor the mean first passage time will be given as follows

\bea{\label{eq:05}}
\left<t(x_{o})\right>=1  &\;\;\;{\mathrm{for}}\;\;\;&  0 < x_{o} \leq 1 \\
\left<t(x_{o})\right>=2  &\;\;\;{\mathrm{for}}\;\;\;& 1 < x_{o} \leq 1+\lambda \nonumber \\
\left<t(x_{o})\right>=3  &\;\;\;{\mathrm{for}}\;\;\;& 1+ \lambda < x_{o} \leq 1+\lambda+\lambda^{2} \nonumber\\
\vdots \nonumber
\eea

\noindent That is, the only walk that could pass from $0 < x_{o} \leq 1$ for the first time
is a walk that takes the first step to the right. Conversely if the walker chooses to go to the left in the first step it falls out of range of $0 \leq x_{o} \leq 1$. The argument also carries to other intervals in (\ref{eq:05}). This result can be put to a more compact form as follows

\be{\label{eq:05a}}
\left<t(x_{o})\right>=\;{\mathrm{int}}\left[\frac{\ln\left[1-(1-\lambda)x_{o}\right]}{\ln\lambda}\;\;\right]\;\;.
\ee

\noindent As can be verified this function increases with unit steps as $x_{o}$ 
reaches certain values given by

\be{\label{eq:05b}}
x_{o}(K)=\frac{1-\lambda^{K}}{1-\lambda}\;\;.
\ee

\noindent These values represent what can be called {\bf extremity walks}. 
That is, walks with all the steps in one direction. 

Thus the mean first passage time has the form of a ladder with step heights increasing by unity and the step
widths decreasing in such a way that they are scaled down with $\lambda$ at each step. In all the cases
we have studied we have observed the ladder pattern, there is an appreciable discrete jump in the mean first passage time as one crosses an extremity walk point. The reason for this pattern is that once the walker reaches a point in say $k$ steps the subsequent walk will have the distribution $P(|x-x_{o}|/\lambda^{k})$. 
However, for general $\lambda$, the step heights and widths of the ladder does not exactly follow
the behavior of the ladder for $\lambda\leq1/2$. Because as one increases $\lambda$ there will be many paths from the origin to a particular $x_{o}$.

We also would like to make a note of the fact that
for $x_{o}=\frac{1}{1-\lambda}-\epsilon$, that is $x_{o}$ being
close to its maximal range with $\epsilon$ small, we get

\be
\left<t(x_{o})\right>=\frac{\ln\epsilon}{\ln\lambda}\;\;.
\ee

\noindent This has the same logarithmic divergence as (\ref{firstas2}) 
if we make the correspondance

\be
\tau\;\;\;\;{\mathrm{is\;\;equivalent\;\;to}}\;\;\;\;-\frac{1}{\ln\lambda}\;\;.
\ee 

\noindent We thus have at least a qualitative connection to the continuum case
as one would expect.

\subsection{Case Study for $\lambda=0.55$}

\begin{figure}[t!]
\includegraphics[scale=0.3]{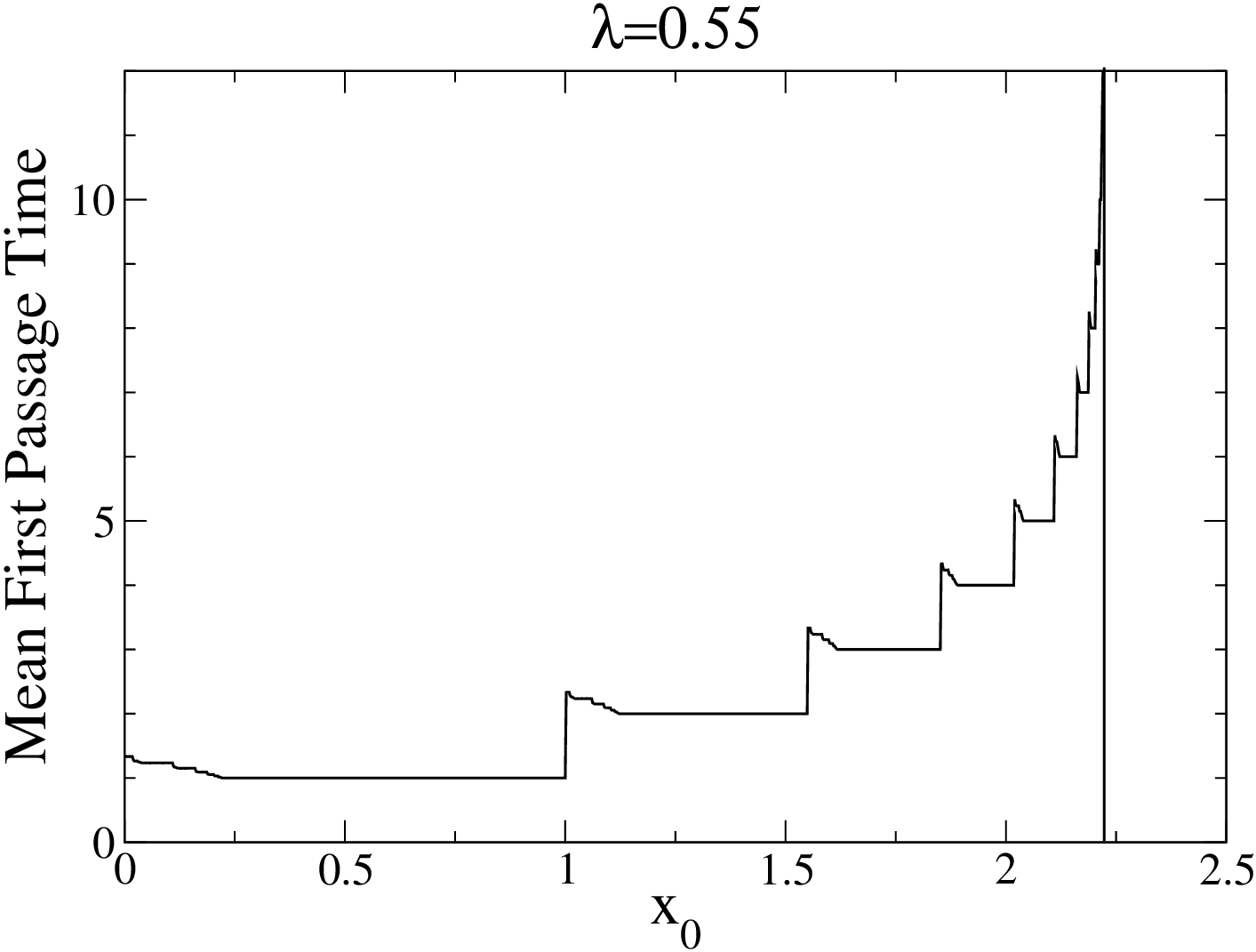}
\caption{{\label{fig:a1}} The mean first passage time from 
$x_{o}$ for $\lambda=0.55$. The vertical line at $x=2.\bar{2}$ 
is the ultimate range $\frac{1}{1-\lambda}$ of the walker for this 
value of $\lambda$. The error bars are very small and are omitted.}
\end{figure}

The result for $\lambda=0.55$ can be seen in Fig.~\ref{fig:a1}.
As we have mentioned, the graph
has a global increase pattern 
and also a self-repetitive, ladder-like local structure. 
That is the function between $x_{o}\in\left[1,1+\lambda\right]$ {\em looks
  like} 
the one between $x_{o}\in\left[0,1\right]$ but scaled with $\lambda$.
This is because once we reach $x_{o}=1$ the subsequent walk can be seen as starting from 
$x_{o}=1$ and with an overall scale of $\lambda$ appearing in front of the distances.
Consequently, to have a {\bf feeling} of the {\em local behavior} 
of the first passage time, {\em it is sufficient to study the graph for mean first passage time for only
$x_{o}\in\left[0,1\right]$}.

\begin{figure}[t!]
\includegraphics[scale=0.3]{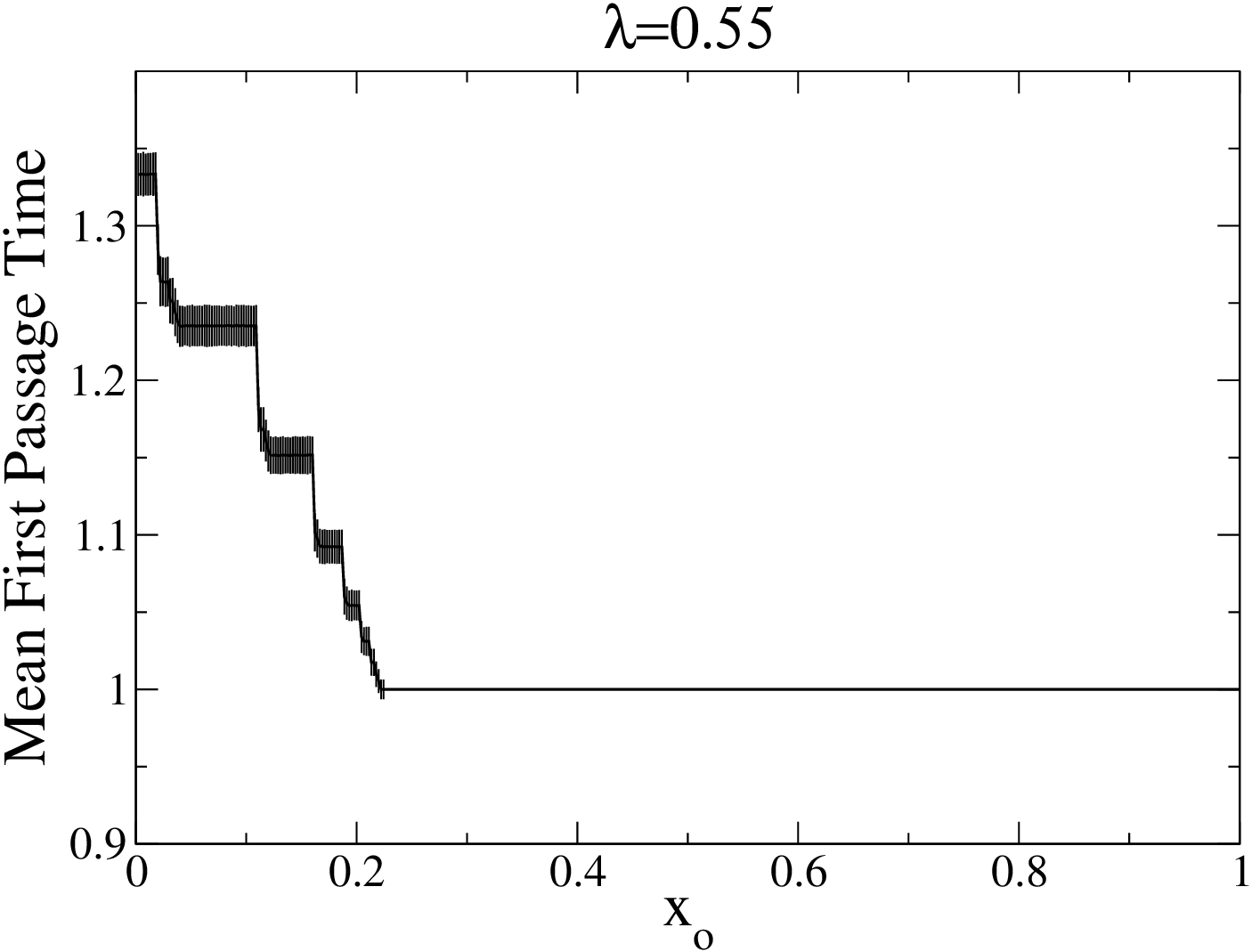}
\caption{{\label{fig:a2}} Zoom of Fig.~\ref{fig:a1} to the 
region $x_{o}\in\left[0,1\right]$. The straight line has zero error. See text for explanation.}
\end{figure}

In Fig.~\ref{fig:a2} which a zoom-out of Fig.~\ref{fig:a1} to 
the region $x_{o}\in\left[0,1\right]$ we see  interesting 
features and we outline their meaning below.

\noindent{\bf 1. Straight line between $x_{o}\in\left[0.\bar{2},1\right]$:} This is a manifestation of the {\em falling out of range} effect we have
outlined above. 
The error bar of this line vanishes and this can be due to only one reason, 
there is one and only one path that can pass from this interval \footnote{One might argue that in a computer simulation a zero error bar can be a fake one if one does not cover the probabilities accurately. This is not the case here and as the text explains the proposition is rigorous.}. 
This becomes easier to understand if we also realize that the value
$0.\bar{2}$ 
is nothing but $-1+\frac{\lambda}{1-\lambda}$ for $\lambda=0.55$. Thus if the walker chooses to
take the first step to the left then the ultimate position 
it can go is simply the value quoted above. Any point beyond 
this value can only be first passed by a single step to the right. 
One could now ask : what should be the largest value 
for $\lambda$ such that there is such a behavior?  
This is answered by the solution of the inequality 

\be{\label{eq:9}}
-1+\frac{\lambda}{1-\lambda} \leq 1 \;\; \rightarrow \lambda \leq \frac{2}{3}\;\;.
\ee  

\begin{figure}[t]
\includegraphics[scale=0.3]{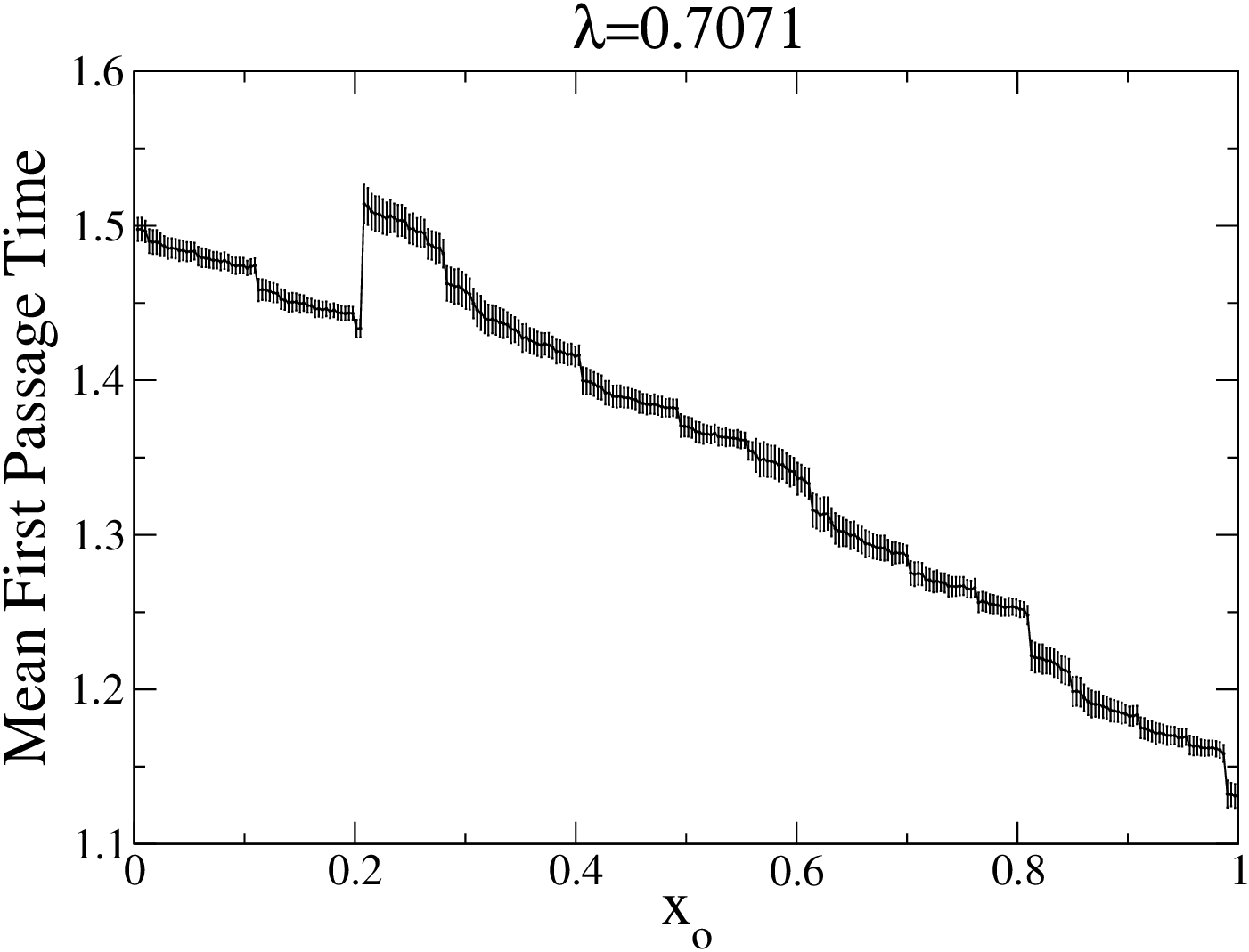}
\caption{{\label{fig:a3}}Mean first passage time for $\lambda=1/\sqrt{2}$ for
the first step of the ladder.}
\end{figure}

\begin{figure}[t]
\includegraphics[scale=0.3]{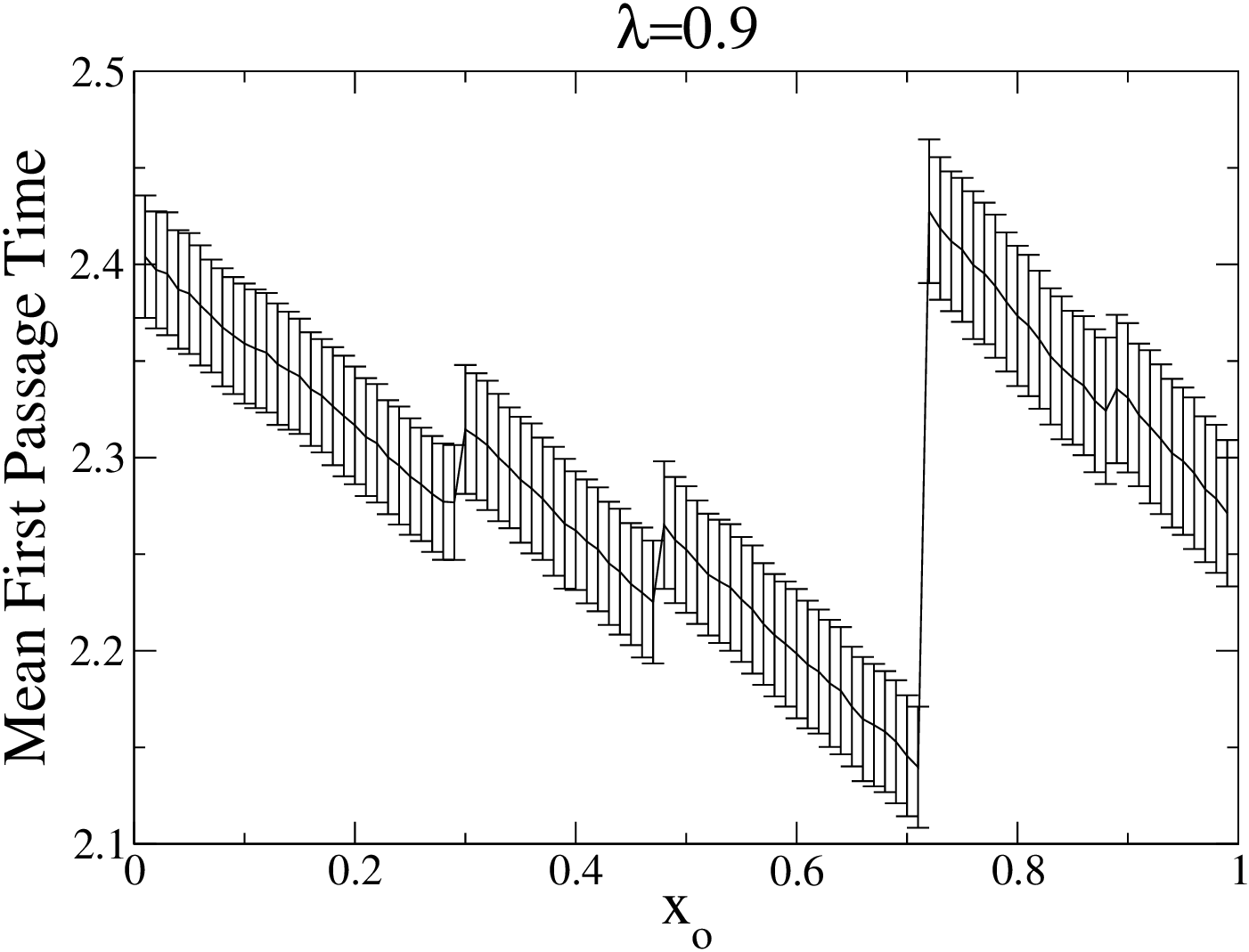}
\caption{{\label{fig:a5}} Mean first passage time for $\lambda=0.9$ for the
  first step of the ladder.}
\end{figure}

\noindent It is instructive to note that this value of $\lambda$ has been observed in \cite{art2} 
to be the smallest value for which there are an infinite number of paths from the origin to any point. 
This, we would like to contrast to the fact that the requirement to have
at least one path from the origin to any point, is satisfied for $\lambda\geq1/2$, again mentioned in \cite{art2}. Here we see an example where the condition is reversed. That is, there are 
only finite number of paths to certain  positions and some positions 
can only be reached by unique paths. For $\lambda\geq 2/3$ we expect 
this straight line behavior to disappear since there will be an infinite number 
of paths from the origin to any point. See Fig.~\ref{fig:a3} for $\lambda=1/\sqrt{3}$ and Fig.~\ref{fig:a5} for $\lambda=0.9$ both of which have $\lambda\geq2/3$.

\noindent{\bf 2. There are multiple plateau's:} This is again a manifestation of the condition that for 
$\lambda\geq 2/3$ there are infinite paths from the origin to any
point. The converse is not necessarily true and it is apparent in Fig.~\ref{fig:a2}. 
Certain values of $x_{o}$  are reached by a {\em finite} (not one) 
number of paths and there are ranges of $x_{o}$ such that this number 
is constant \footnote{Since the number of paths  is not "one" we have error bars for these cases.}. Consequently, the appearance of plateau's is guaranteed by 
the condition $\lambda\leq 2/3$ and we see this to disappear in Fig.~\ref{fig:a3} and Fig.~\ref{fig:a5} (both of which represent cases for $\lambda\geq 2/3$) where the plateau structure
has transformed into a cascaded hill form.

\noindent{\bf 3. Time goes up as $x_{o}$ gets small?}  If we consider Fig.~\ref{fig:a1} and Fig.~\ref{fig:a2},
we observe that as opposed to the global increase of the ladder, the mean first passage time has a local increase pattern if we go to low $x_{o}$. This local increase, as we 
have outlined above, is due to the fact that one particular path might 
contribute to the first passage distribution of more than one $x_{o}$, and as $x_{o}\to 0$ the number of paths that do so generally increase without decreasing the probability of the first passage distribution. 
On the other hand we expect as $\lambda\to 1$ that this 
behavior will disappear since the number of $x_{o}$'s, a particular 
path might contribute, would naturally decrease.

\begin{figure}[t]
\includegraphics[scale=0.3]{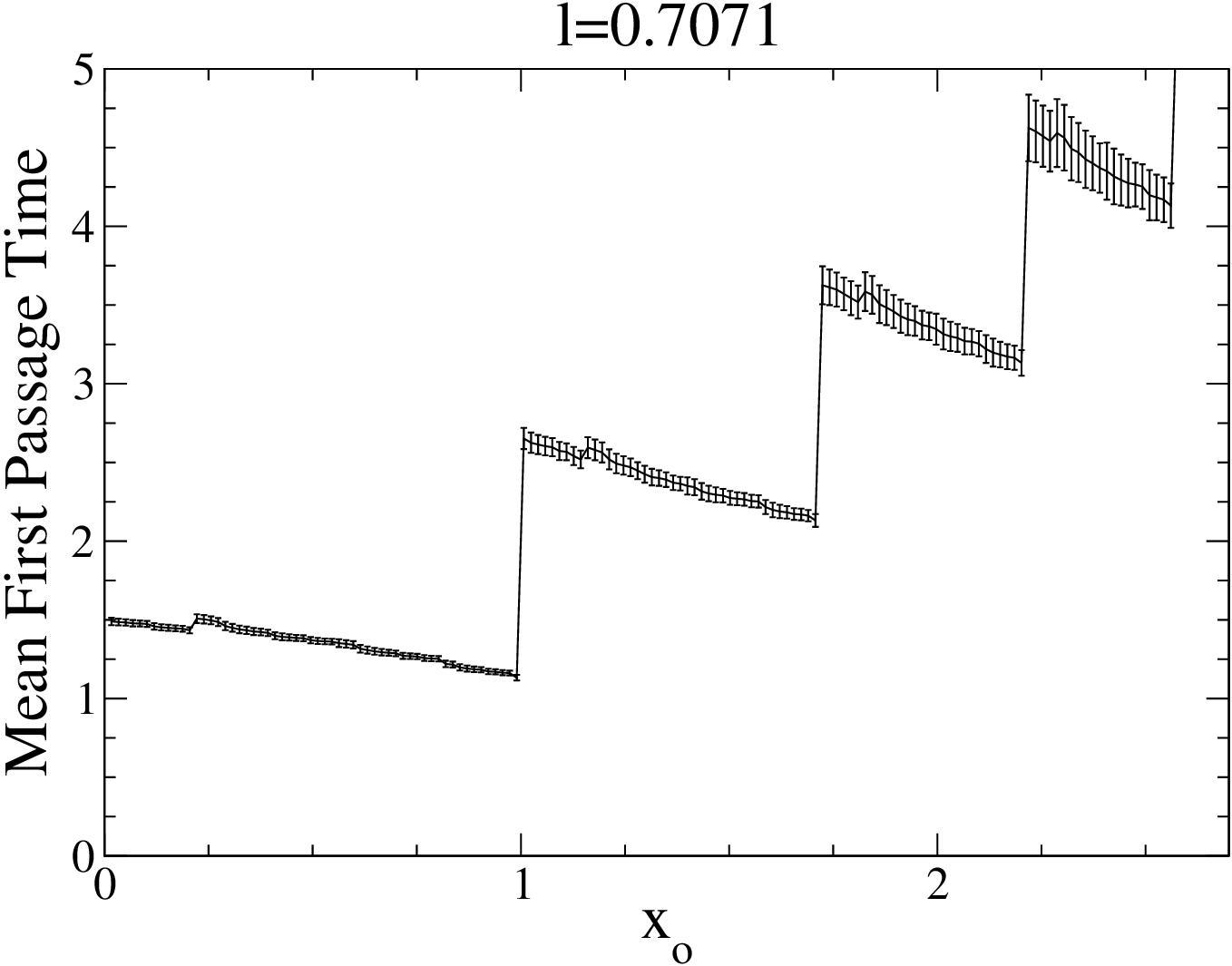}
\caption{{\label{fig:more}}Mean first passage time for $\lambda=1/\sqrt{2}$ for
the first few steps in the ladder.}
\end{figure}

\subsection{Intermediate Conclusions}

Here we would like to sum up what we have observed so far. For each $\lambda$ the curve for the 
mean first passage time has the form of a ladder. The ladder has appreciable jumps at the extremity walk
points $x_{o}(K)$ given by (\ref{eq:05b}). These jumps are forming the backbone of the global increase as $x_{o}$ gets larger. However on a particular step of the ladder, that is for $x_{o}(K)\leq x_{o}\leq x_{o}(K+1)$, there is very rich local behavior. We have studied some of these. On the other hand, this local behavior is not very fierce. For example, if we consider the first step of the ladder for different $\lambda$'s as in Figs.~\ref{fig:a2}, \ref{fig:a3} and \ref{fig:a5}, we see that the mean first passage time is not too different than the value to the rightmost value of the step. Therefor to study the global increase
pattern it should be enough to consider only the first passage time from  extremity walk points $x_{o}(K)$.

\section{Global Structure for  General $\lambda$}\label{sec5}

For $\lambda\geq 2/3$ the mean first passage time differs from (\ref{eq:05a}) 
for the extremity walks defined in (\ref{eq:05b}) 
as can 
be checked from Fig.~\ref{fig:a3} and Fig.~\ref{fig:a5} where 
the mean first passage times are not 1 for $x_{o}(1)=1$. Nevertheless,
as we have mentioned, with increasing $x_{o}$  the ladder structure is still present, and the  ladder 
jumps are still occuring at $x_{o}(K)$. So we have to consider another approach.
To study how much discrepancy arises, we have simulated the random 
walk for different extremity points   and computed 
the mean first passage time minus the
values that is given by (\ref{eq:05a})
, that is for each $\lambda$ we computed 

\be
\left<t^{\prime}(K)\right>\equiv\left<t(x_{o}(K))\right>-K\;\; .
\ee

\begin{figure}[tp]
\includegraphics[scale=0.3]{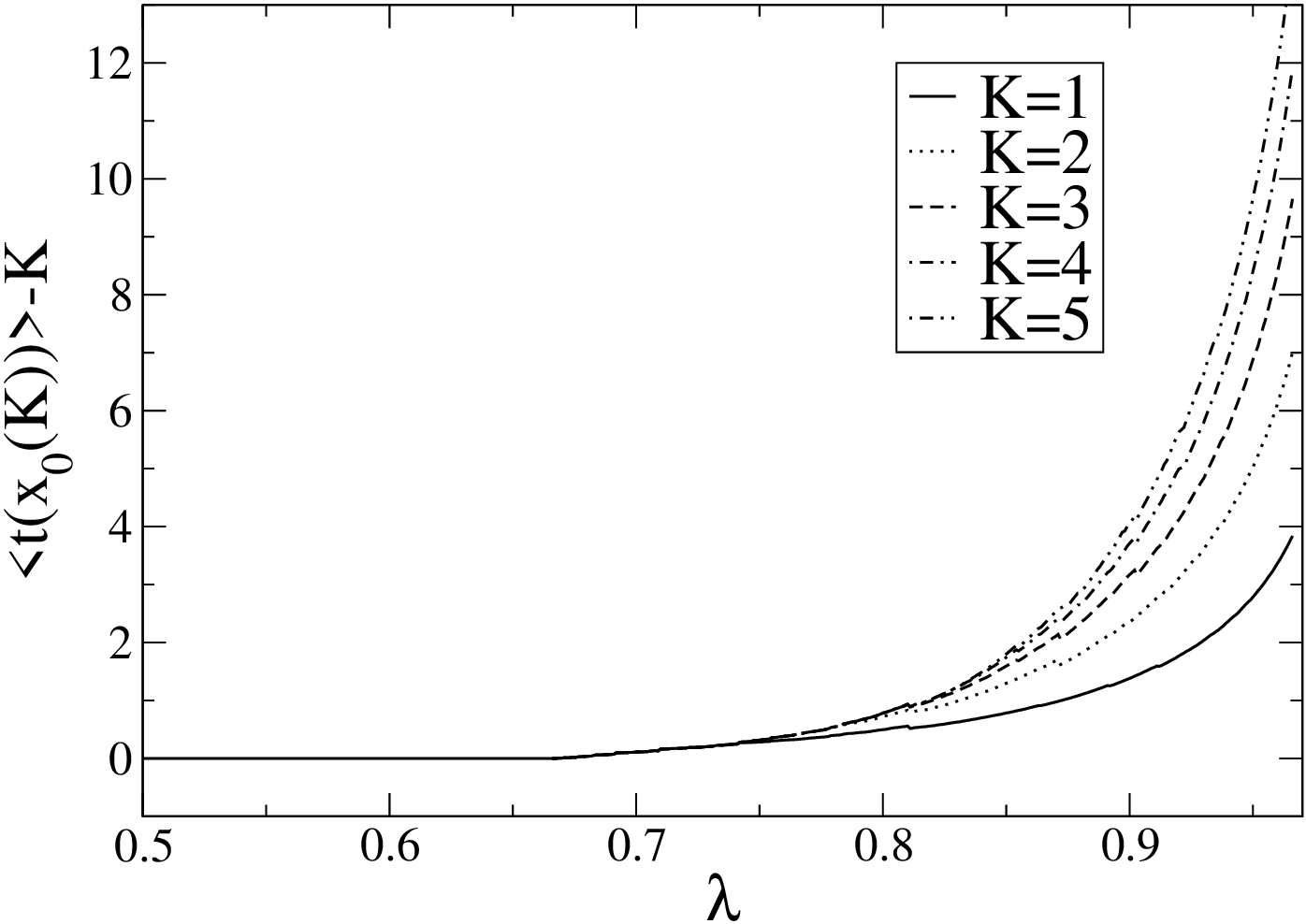}
\caption{{\label{fig:a6}} $\lambda$ behavior of the first passage time for various extremity walk points. Error bars are omitted for clarity}
\end{figure}

\begin{figure}[t]
\includegraphics[scale=0.3]{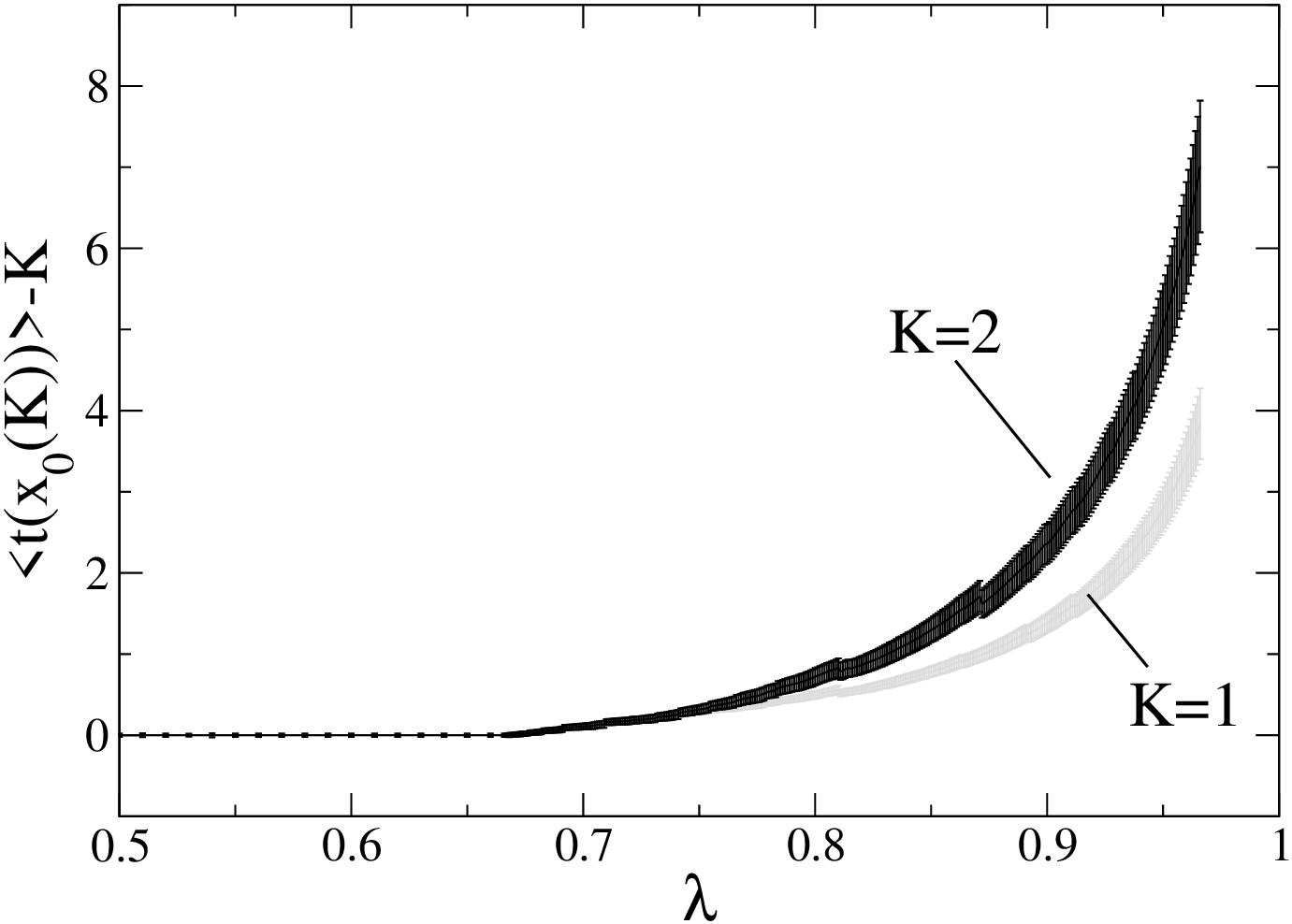}
\caption{{\label{fig:a7}} Fig.~\ref{fig:a6} shown with error 
bars for the first two extremity walk points.}
\end{figure}

\noindent The result is presented in
Fig.~\ref{fig:a6}, where we have again suppressed the error bars to have a clear picture. 
A representative picture with the error bars are given in Fig.~\ref{fig:a7}. 

The discrepancy starts at $\lambda=2/3$ as anticipated. It is small for 
not too large values of $\lambda$ and diverges later on. 
However we see that the discrepancy itself has a pattern.
All the curves agree up to a certain value of $\lambda$, 
then the curve for $K=1$ starts to deviate. After this point the 
remaining ones
agree up to a another certain value of $\lambda$ after which the curve for $K=2$ 
starts to deviate. This goes on in the same way for higher values of $K$. 

To understand and quantify this behavior we would like to remind 
the reader that extremity walks are in essence the fastest paths to a 
given point. Then, the next to fastest path would be
one with a single  step in the wrong direction.  
Of course it is important  when the wrong step is realized and 
this will be the crucial point in our line of reasoning. 
Let us try to construct our argument by an example: consider a first 
passage path from the origin to
$x_{o}(2)$, the fastest path is of course an extremity walk that 
would take two steps to the right. 
Next, we would like to consider paths with one 
step in the wrong direction that would still  pass from 
$x_{o}(2)$. There are two possibilities: one can either 
choose the wrong direction in the first step or in the second. 
Considering these choices and demanding the subsequent walk 
would be in range, we have two equations

\bea
-1+\frac{\lambda}{1-\lambda}&=&1+\lambda \rightarrow \lambda\simeq0.732051\label{eq:birinci}\;\; ,\\
1-\lambda+\frac{\lambda^{2}}{1-\lambda}&=&1+\lambda \rightarrow \lambda=\frac{2}{3}\label{eq:ikinci}\;\; .
\eea

Now (\ref{eq:ikinci}) is the same as the condition to reach $x_{o}(1)$ 
with one step in the wrong
direction, so it is not a new condition. However (\ref{eq:birinci}) 
is new and once $\lambda$ gets bigger than the value quoted this 
walk (which was already contributing to $x_{o}(1)$) 
will start to contribute to the first passage time for $x_{o}(2)$.
Thus between these two values of $\lambda$, $\left<t^{\prime}(K)\right>$ will  be the same
for $K=1$ and $K=2$, and consequently for all $K$.
However as soon as $\lambda$ gets bigger than $0.732051$ a new path 
will start to contribute to the first passage time for $x_{o}(2)$,
meaning that its difference from the base value $2$ will be bigger. 
This is why the curves for $K=1$ and the rest start to branch at 
this point.

To iterate and make the idea clearer let us know consider 
the same argument for $x_{o}(3)$. 
There are three possible ways one can opt for the wrong 
direction and demanding the walks will be in range we get the following

\bea
-1+\frac{\lambda}{1-\lambda}&=&1+\lambda+\lambda^{2} \label{eq:ucuncu}\;\; , \\
1-\lambda+\frac{\lambda^{2}}{1-\lambda}&=&1+\lambda+\lambda^{2} \;\; ,  \\
1+\lambda-\lambda^{2}+\frac{\lambda^{3}}{1-\lambda}&=&1+\lambda+\lambda^{2} \;\; ,
\eea

\noindent which would yield respectively,

\bea
\lambda&\simeq&0.770917 \label{eq:ucuncu2}\;\; , \\
\lambda&\simeq&0.732051\;\; , \\
\lambda&=&\frac{2}{3}\;\; .
\eea

\noindent We have indeed numerically observed the  fact that 
the curves for $K=2$ and $K=3$ agree up to around $\lambda\simeq 0.770917$.
Since as $\lambda$ gets bigger than this value a path that was already
contributing to $x_{o}(1)$ and $x_{o}(2)$ would be added 
to the  path list of $x_{o}(3)$ and hence creating a shift in the curve.

It is also possible to argue about the existence of paths with 
two or more steps in the wrong direction which will in general
give an infinite number of possible constraint equations and 
possibly interfering with the picture above. Let us consider 
the case with two steps in the wrong direction. There are three 
generally possible cases. First, one could choose the wrong 
steps early in the walk and not so separated in time thus subjecting 
the subsequent walk to greater punishment. Second, one could choose 
them after having subsequent steps to the right, again not so separated in
time, meaning a much less punishment for the subsequent walk. 
Or third, one could have them separated by a long walk to the right for which the punishment 
would be somewhere in between compared to the other two cases. 
It is clear that only the first case comes with more probability because 
less number of steps are fixed \footnote{This line of reasoning about the
 probability of paths might seem speculative at first  
since we actually let the walker take an infinite steps to the right after 
all the wrong steps are taken. However this is only a constraint equation, 
once $\lambda$ gets bigger than the value given by the constraint, 
the last phase of the walk will certainly consist of a finite number of 
steps with step number decreasing with increasing $\lambda$. But the 
number of steps fixed at the beginning would still be reducing the overall 
probability of the path in question.}. So we would like to introduce the 
idea of {\bf worst case punishment extremity walk} defined as the path in which all the
wrong steps are taken at the beginning of the walk. This is a 
generalization of the idea we outlined above. The general constraint 
equation for the $P$-step worst case punishment extremity walk (that is $P$ steps 
to the left at the very beginning) to go to $x_{o}(K)$ extremity point is given by

\be
\lambda^{K}+2\lambda^{P}-2=0\;\;.
\ee

\noindent the possible roots of this equation for the 
first few values of $(K,P)$ and are presented below

{\begin{center}
\begin{tabular}{|c||c|c|c|}
\hline 
P $\setminus$ K & 1 & 2 & 3 \\
\hline\hline
1 & 0.666667 & 0.732051 & 0.770917  \\
\hline
2 & 0.780776 & 0.816497 & 0.839287 \\
\hline
3 & 0.835122 & 0.858094 & 0.87358  \\
\hline
\end{tabular} 
\end{center}}

From this table the fourth highest value is given 
for $(K=1,P=2)$ and not $(3,1)$. This means that just 
before the path $(3,1)$ contributes to $x_{o}(3)$ the 
path $(1,2)$ starts to contribute to $x_{o}(1)$ and this 
condition would also satisfy a different walk of the form
$1-\lambda-\lambda^{2}+\frac{\lambda^{3}}{1-\lambda}=x_{o}(2)=1+\lambda$. 
So the process of branching we have mentioned above is actually more complex. 
However the behavior from $\lambda=0.66666$ to $\lambda=0.732051$ is 
free of these complexities. Thus one can safely state that we have the
following

\be
\left<t^{\prime}(K)\right>=f(\lambda)\;\;\;\; {\rm for} \;\;
\frac{2}{3}\leq\lambda\leq0.732051 \;\; ,
\ee

\noindent a function independent of $K$. 

As a final remark we would like to mention that all the curves in Fig.~\ref{fig:a6} fit
very well to the following ansatz

\be
\frac{a_{0}(K)}{\sqrt{-\ln\lambda}}+a_{1}(K)+\frac{a_{2}(K)}{\ln\lambda}\;\;,
\ee

\noindent with $a_{0}$, $a_{1}$ and $a_{2}$ are fit parameters. 
Here the first two terms, which dominate the small $\lambda$ behavior, represent an analogy the 
small $u$ behavior 
of (\ref{firstas}), whereas the last one, which dominate the large $\lambda$ behavior, is 
the analog of large $u$ asymptotic behavior in (\ref{firstas2}), if we would like to make the
qualitative identification $\tau\sim-1/\ln\lambda$. 
Since, with this  identification and for fixed
$x_{o}(K)$, increasing (decreasing) $\lambda$ would be similar to increasing
(decreasing) $u$. So the discrete case is not
qualitatively very different than the continuum model we have
presented.

\section{Conclusion}

In this work we have studied the first passage characteristics of a random
walker with step sizes decaying exponentially in time. There are
rich mathematical structures which we have studied via
computer simulations. We have also shown that the discrete case
shares all the qualitative properties of a diffusion equation with an exponentially decaying diffusion constant which
hints phenomenologically that the continuum case might be of choice for studying general aspects of such physical systems.

Although, mainly to connect to the literature, we have confined ourselves to the 
exponentially decaying step sizes, there are many other possibilities. One possibly interesting
example that comes to  
mind is to assume $s_{N}=\left(1-\frac{N}{M}\right)^{\alpha}$ with $M$ integer and $\alpha$ arbitrary. 
This case has the nice extra feature that the walk really comes to a stop when the walker commits $M$ 
steps therefor allowing an exact enumeration of paths for small values of $M$. Furthermore this 
example has a connection to the exponential case remembering $e^{x}=\lim_{a\to\infty}(1+\frac{x}{a})^a$. A 
preliminary analysis we have carried out gives similar behavior to the exponential case while differing 
in details. Another likely extension is, as proposed in \cite{art1}, the n-dimensional random walker 
with shrinking steps which might provide further interesting features.

\begin{acknowledgments}
The computer simulations for this work have been carried out on the 16-node Gilgamesh PC-cluster at
Feza G\"ursey Institute. We would like to thank 
A. Erzan,  M. Mungan and E. Demiralp for useful discussions and comments on this work.

\end{acknowledgments}



\begin{thebibliography}{1}
\bibitem{art1} P.L. Krapivsky and S. Redner, Am.J.Phys. {\bf 72}, 591 (2004). 
\bibitem{art2} A.C. de la Torre, A. Maltz, H.O.M\'artin, P. Catuogno and
  I. Garci\'a-Mata, Phys.Rev.E {\bf62}, 7748-7754 (2000).
\bibitem{art3} E. Barkai and R. Silbey, Chem. Phys. Lett. {\bf 310}, 287-295 (1999).
\bibitem{art4} E. Ben-Naim, S. Redner and D. ben-Avraham, Phys. Rev. A {\bf 45}, 7207-7213 (1992).
\bibitem{art5} B. Jessen and A. Wintner, ``Distribution functions and the
  Riemann zeta function'', Trans.\ Amer.\ Math.\ Soc.\ {\bf 38}, 48-88
  (1935); B. Kershner and A. Wintner, ``On symmetric Bernoulli
  convolutions'', Amer.\ J. Math.\ {\bf 57}, 541-548 (1935); A.  Wintner,
  ``On convergent Poisson convolutions'', {\it ibid.} {\bf 57}, 827-838 (1935).
\bibitem{art6} P. Erd\H os, ``On a family of symmetric Bernoulli convolutions'',  Amer. J. Math. {\bf 61}, 974-976 (1939); P. Erd\H os, ``On smoothness
  properties of a family of Bernoulli convolutions'', {\it ibid.} {\bf 62},
  180-186 (1940).

\bibitem{art7} A. M. Garsia, ``Arithmetic properties of Bernoulli
  convolutions'', Trans.\ Amer.\ Math.\ Soc.\ {\bf 102}, 409-432 (1962); A. M.
  Garsia, ``Entropy and singularity of infinite convolutions'', Pacific J.
  Math.\ {\bf 13}, 1159-1169 (1963).

\bibitem{art8}  J. C. Alexander and J. A. Yorke, ``Fat baker's transformations'', Ergodic Th.\ Dynam.\ Syst.\ {\bf 4}, 1-23 (1984); J. C. Alexander and D.
  Zagier, ``The entropy of a certain infinitely convolved Bernoulli
  measure'', J. London Math.\ Soc.\ {\bf 44}, 121-134 (1991).
\bibitem{art9} F. Ledrappier, ``On the dimension of some graphs'', Contemp.\
  Math.\ {\bf 135}, 285-293 (1992).
\bibitem{art10} Y. Peres, W. Schlag, and B. Solomyak, ``Sixty years of
  Bernoulli convolutions'', in {\it Fractals and Stochastics II}, edited by
  C. Bandt, S. Graf, and M. Z\"ahle (Progress in Probability, Birkhauser,
  2000), Vol. 46, pp. 39-65.
\bibitem{art11} S. Redner, ``A Guide to First-Passage Processes'', Cambridge University Press, 2001.
\bibitem{art12} G. Rangarajan and M. Ding, Phys. Lett. A {\bf 273}, 322 (2000); G. Rangarajan and M. Ding, Phys. Rev. E {\bf 62} 120 (2000); G. Rangarajan and M. Ding, Fractals {\bf 8}, 139 (2000).
\bibitem{art13} M. Gitterman, Phys. Rev. E {\bf 62}, 6065 (2000).
\bibitem{art14} R. Metzler, Phys. Rev. E {\bf 63}, 012103 (2000).
\bibitem{art15} E. Barkai, Phys. Rev. E {\bf 63}, 046118 (2001).
\bibitem{art16} K.S. Fa and E.K. Lenzi, Phys. Rev. E {\bf 67} 061105 (2003).
\end{thebibliography}
\end{document}